# Imaging the dynamics of cardiac fiber orientation in vivo using 3D Ultrasound Backscatter Tensor Imaging

Clement Papadacci[1§], Victor Finel[1§], Jean Provost[1], Olivier Villemain[1], Patrick Bruneval[2], Jean-Luc Gennisson[1], Mickael Tanter[1], Mathias Fink[1], Mathieu Pernot[1*]

[1]Institut Langevin, ESPCI ParisTech, CNRS UMR 7587, INSERM U979, 17 rue Moreau, 75012 Paris, France

[2]Department of Pathology, Hôpital Européen Georges Pompidou, 21, rue Leblanc 75015, Paris,

§Co-first authors: both authors contributed equally to this work

*Corresponding author: Mathieu Pernot, Institut Langevin, 17 rue Moreau, 75012 Paris, +33 1 80 96 30 40, mathieu.pernot@inserm.fr

**Abstract**

The assessment of myocardial fiber disarray is of major interest for the study of the progression of myocardial disease. However, time-resolved imaging of the myocardial structure remains unavailable in clinical practice. In this study, we introduce 3D Backscatter Tensor Imaging (3D-BTI), an entirely novel ultrasound-based imaging technique that can map the myocardial fibers orientation and its dynamics with a temporal resolution of 10ms during a single cardiac cycle, non-invasively and *in vivo* in entire volumes. 3D-BTI is based on ultrafast volumetric ultrasound acquisitions, which are used to quantify the spatial coherence of backscattered echoes at each point of the volume. The capability of 3D-BTI to map the fibers orientation was evaluated in vitro in 5 myocardial samples. The helicoidal transmural variation of fiber angles was in good agreement with the one obtained by histological analysis. 3D-BTI was then performed to map the fiber orientation dynamics in vivo in the beating heart of an open-chest sheep at a volume rate of 90 volumes/s. Finally, the clinical feasibility of 3D-BTI was shown on a healthy volunteer. These initial results indicate that 3D-BTI could become a fully non-invasive technique to assess myocardial disarray at the bedside of patients.

**Introduction**

The myocardial fibers architecture plays a major role in the cardiac function. Fibers orientations vary continuously and smoothly through myocardial walls[1] and this complex organization is closely linked to the mechanical and electrical myocardial function[2–4]. For instance, the helicoidal fiber distribution in the left ventricular walls contributes to the torsional motion of the heart during the ejection phase, which is believed to improve the heart's pumping efficiency[5,6]. The fibers of the heart are also related to its electromechanical properties[7–10]; indeed, the electrical activation propagates preferentially in the direction of myocardial fibers and results in their synchronous contraction. Myocardial fiber disarray is thought to appear in the early stage of many pathologies such as in cardiomyopathies or in fibrosis[11]. The mapping of the fiber architecture, on the one hand, could increase our knowledge of the cardiac function[12], and on the other hand, could potentially enable early diagnosis of cardiomyopathies. Yet, no method to map the fiber architecture is currently used on a regular basis for clinical purposes.

Magnetic Resonance Diffusion Tensor Imaging (MR-DTI)[13,14] is widely used for mapping the structural connectivity of the human brain. However, its application to the heart, while feasible in animals[15] and humans[16], remains challenging in vivo due to limited frame rates (and hence long acquisition times) and limited robustness of the MR-DTI signals to tissue motion[11,15,16]. Optical methods such as optical coherence tomography[17] and two-photon microtomy[18] can also map the fiber directions at the microscopic level but remain limited to superficial *ex vivo* tissue and to small regions of interest. In ultrasound imaging, several methods have been proposed to quantify the anisotropy of various physical parameters linked to the fiber orientation, including the ultrasonic attenuation[19,20], the integrated backscattered intensity[19–22] and the myocardial stiffness using Elastic Tensor Imaging (ETI)[23]. However, none of these methods has been yet implemented for the time-resolved, 3D mapping of the myocardial fibers orientation.

In this paper, an entirely novel ultrasound-based imaging technique called Backscatter Tensor Imaging (3D-BTI) is described and evaluated in vivo with respect to its capability of mapping the myocardial fibers orientation and its time-resolved dynamics during an entire cardiac cycle. 3D-BTI is based on ultrafast volumetric ultrasound acquisitions, which are used to quantify the spatial coherence of backscattered echoes at each point of the imaged volume at a rate of 90 volumes/s. Spatial coherence is a physical property of backscattered waves that depends on the distribution of scatterers at the subwavelength scale. Ultrasonic spatial coherence has thus the capability to provide statistical information on the tissue microstructure at a scale far below the resolution of ultrasound images. In 3D-BTI, the anisotropy of spatial coherence is analyzed on a 2D matrix probe in order to derive its principal directions and derive the fibers orientation. The link between the anisotropy of spatial coherence and the fibers orientation was previously demonstrated in composite materials and in fibrous soft tissues such as the myocardium and the skeletal muscle[24,25].

Another key aspect of 3D-BTI is the use of volumetric ultrafast plane wave imaging. We recently developed a 3D fully programmable ultrasound system for volumetric ultrafast plane wave imaging achieving volume rates of up to 5000 volumes/s[26]. By coherent compounding of the backscattered echoes associated with tilted plane wave emissions, the spatial coherence and thus the fibers orientation can be obtained simultaneously in each voxel of the volume at high volume rate.

3D-BTI is hereby found to be capable of mapping the transmural fibers orientation in a beating heart. For the purpose of this validation study, 3D-BTI was compared to histological analysis on ex vivo explanted myocardial tissues. An excellent agreement on the transmural fiber orientation was found between 3D-BTI and histology. The in vivo feasibility of 3D-BTI on a beating heart was then demonstrated in an open chest ovine model. Finally, transthoracic imaging was performed on a healthy volunteer to show the clinical feasibility of 3D-BTI. To our knowledge, it is the first time that the dynamics of the myocardial fibers are observed non-invasively over an entire cardiac cycle of a single heartbeat in humans.

**Results**

The general principle of 3D-BTI is illustrated in figure 1. First, tilted plane waves (figure 1 A) are emitted from a 2D matrix array probe connected to a customized, programmable, ultrasound system[26]. For each emitted plane wave, the backscattered echoes received by each element of the matrix probe are recorded and further processed using 3D plane-wave coherent compounding to synthetically generate a voxel-specific focal region (figure 1 B). The spatial coherence associated to each focal region is then computed at each point of the 3D volume (figure 1 C) and used to determine the fiber orientation by applying an elliptic fit (figure 1 D). Finally, a vector representation is used for the visualization of fibers in 3D space (i.e. figure 1 E).

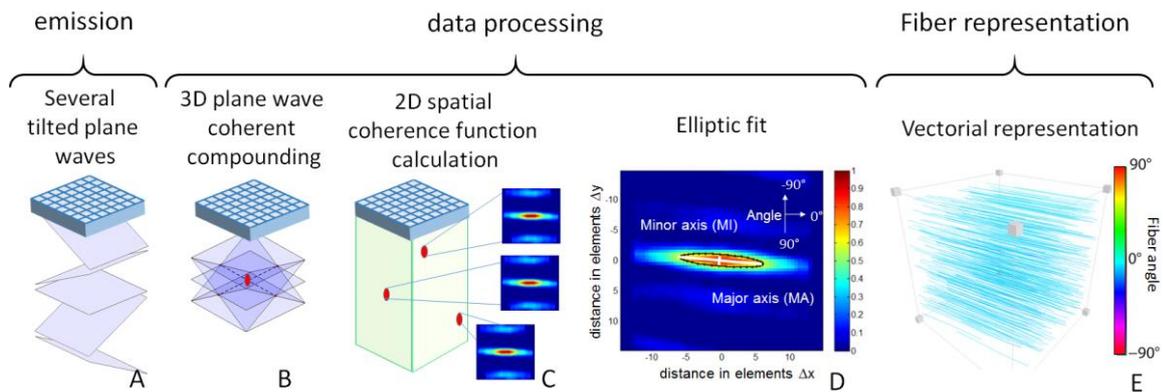

*figure 1* ***3D-BTI principle***. *A. Tilted plane waves are emitted and the associated backscattered echoes are recorded. B. Coherent compounding is applied in postprocessing to synthetically generate voxel-specific focal zones. C. The 2D spatial coherence function is calculated for each voxel and D. an elliptic fit is applied to determine the fibers orientation. E. a vector representation is then achieved to display the fibers orientation in 3D.*

*Ex vivo experiments*

To establish the accuracy and precision of 3D-BTI, myocardial samples explanted from the anterior wall of the left ventricle of a porcine heart were imaged using 3D-BTI and further analyzed by histology. figure 2.A shows the 3D grayscale ultrasound image of one myocardial sample and three examples of spatial coherence functions obtained at different depths (figure 2 B, C, D). At each location, the spatial coherence presents a strong anisotropy with a principal direction that varies with depth. The vector representation (figure 2.F) of the entire volume shows that fiber orientation varies across the wall. Fiber orientation was found to vary gradually through the ventricular wall, which is consistent with the literature[1]. The transmural orientation was averaged over the volume for the 5 myocardial samples and was found to vary continuously through the wall with an average difference of 98.6°± 8.9° between endocardium and epicardium. These results were compared against the fiber orientation obtained from histological analysis. Figure 3.A and B shows a representative sample studied with both histology and 3D-BTI. Figure 3.C shows a Bland-Altman analysis applied to all five samples of this study. A bias of 0.33 degrees and 95-% limits of agreement equal to -10.1° and 10.7° were found.

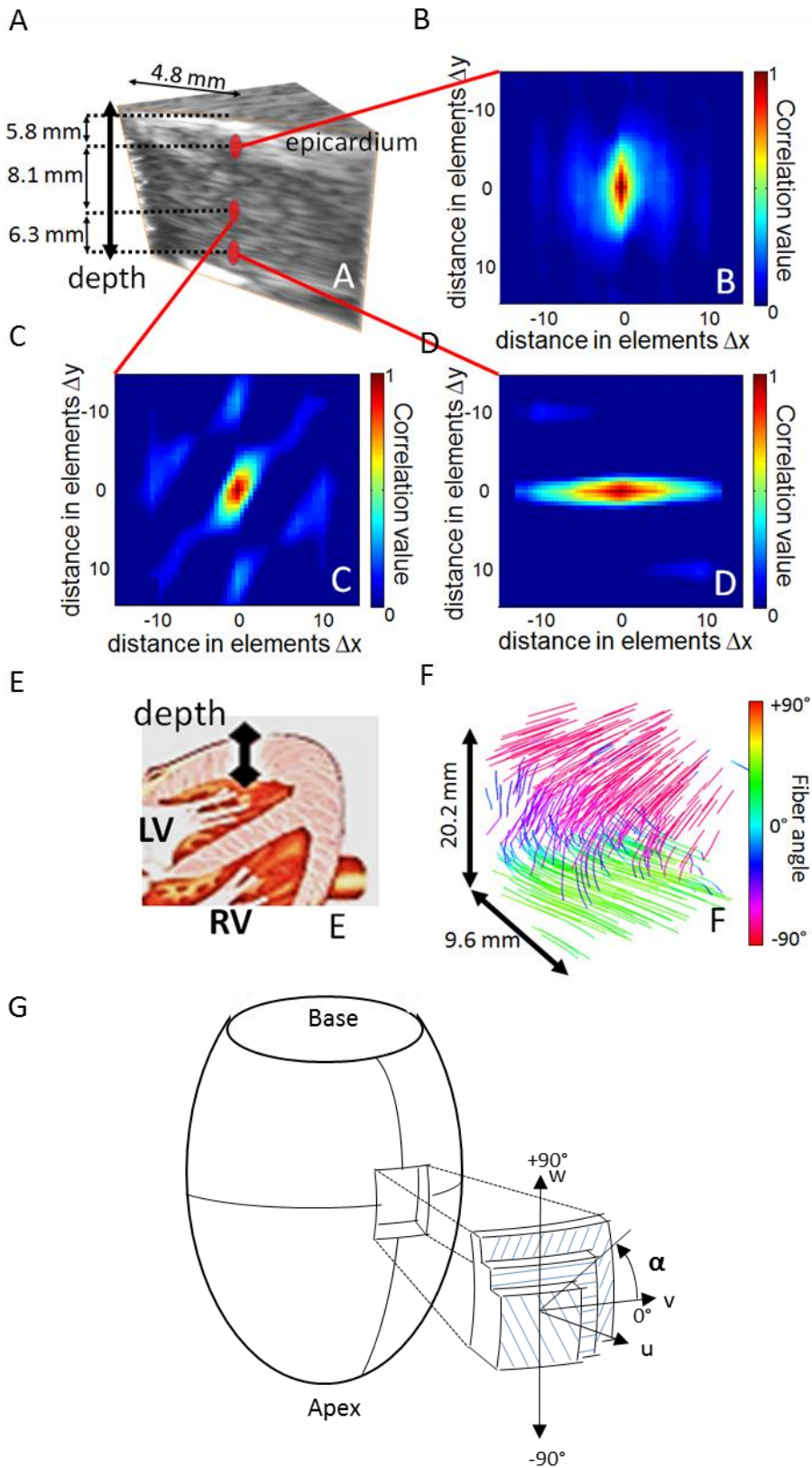

figure 2    **3D-BTI of one ex vivo porcine myocardial sample**. A 3D grayscale ultrasound image of one myocardial sample. B, C, D Three examples of spatial coherence functions

*obtained for different depth locations in the sample. E drawing of the heart associated to the ultrasound acquisition location. F Vector representation used for the visualization of the transmural variation of the fibers in the entire sample. G **Schematic representation of the reference coordinates.** Local axes u, v and w are defined by the circumferential and longitudinal directions on the epicardial surface. Fiber angle α, is positive when measured counterclockwise from the v-axis. (Figure adapted from Streeter et al.* [27]*)*

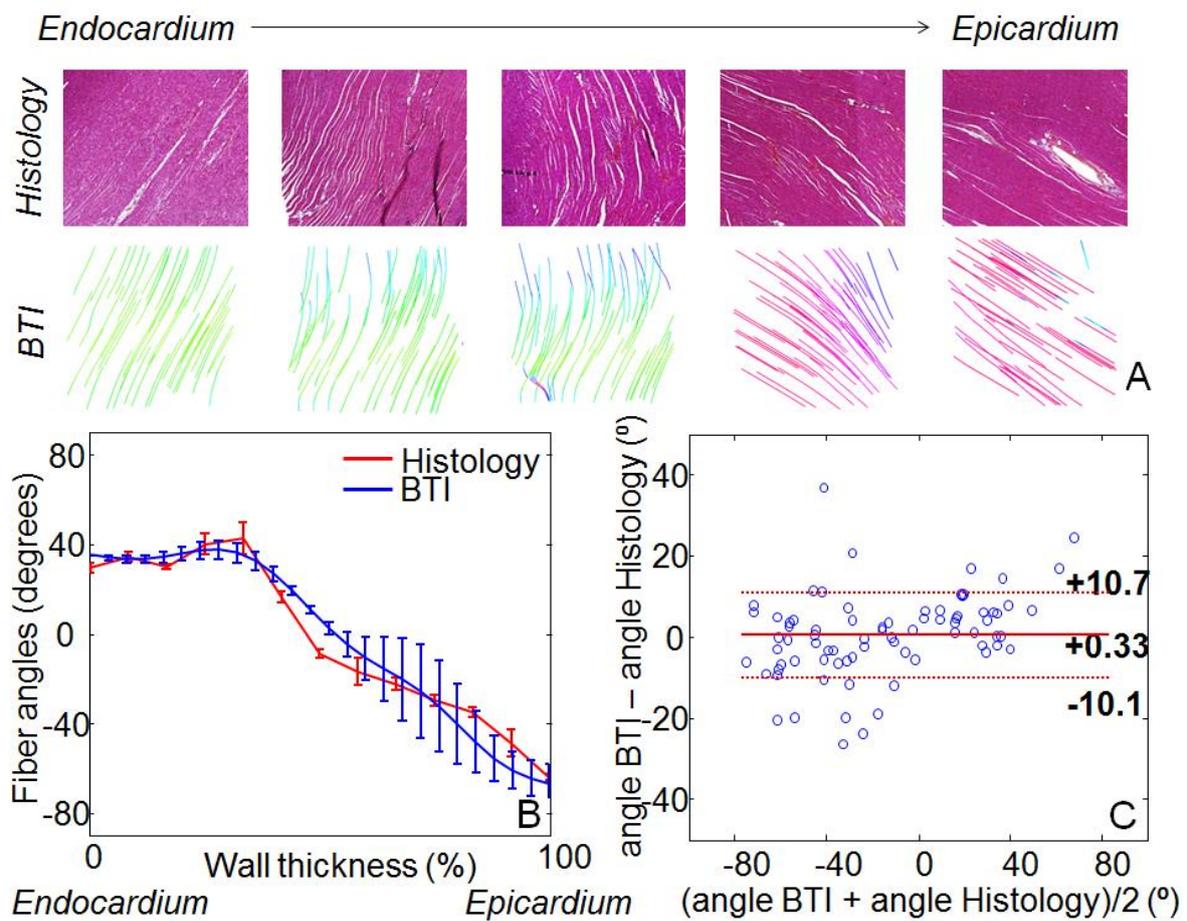

figure 3    ***3D-BTI validation against histology***. *A. Five histological slices (20X) and associated BTI slices of a left ventricle myocardial sample highlights the transmural angle variation of the fibers. B. An example of the fiber angles variation assessed with histology and with 3D-BTI through the wall thickness (0% endocardium -100% Epicardium) for the same*

*sample is displayed. C. Bland-Altman plot of the transmural fiber angles estimated by 3D-BTI and histology (n=5 myocardium samples).*

*In vivo open chest experiments*

The in vivo feasibility of 3D-BTI was demonstrated in the beating heart of an open chest sheep. 3D-BTI acquisitions were performed during a complete single cardiac cycle with a volume rate of 90 volumes/s. Figure 4 shows examples of fiber reconstruction at specific times of the cardiac cycle (A Late Diastole, B Early Systole, C Late Systole, D Early Diastole). It demonstrates the feasibility of following the fiber orientation during an entire cardiac cycle at high frame rate (the complete cine-loop is shown in supplemental movie 1). The transmural fiber distribution at one location as a function of time is shown in figure 5. Fiber angles were found to vary as a function of depth with an absolute difference of 96° from the epicardium to the endocardium.

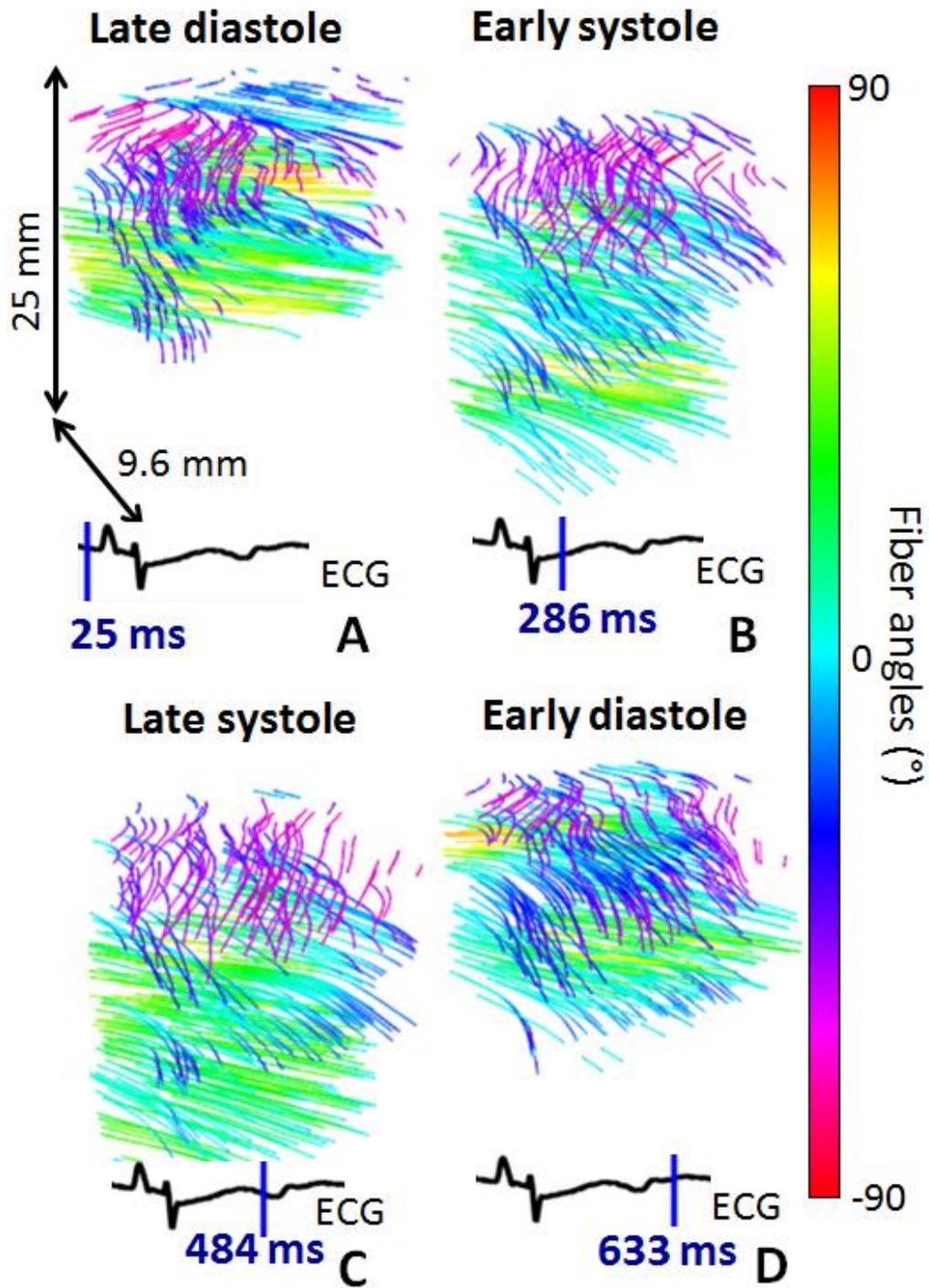

*figure 4* **In vivo 3D-BTI.** *3D representation of fibers orientation in the left ventricle of an open-chest sheep at four different moments of the cardiac cycle (i.e. A Late Diastole, B Early Systole, C Late Systole, D Early Diastole).*

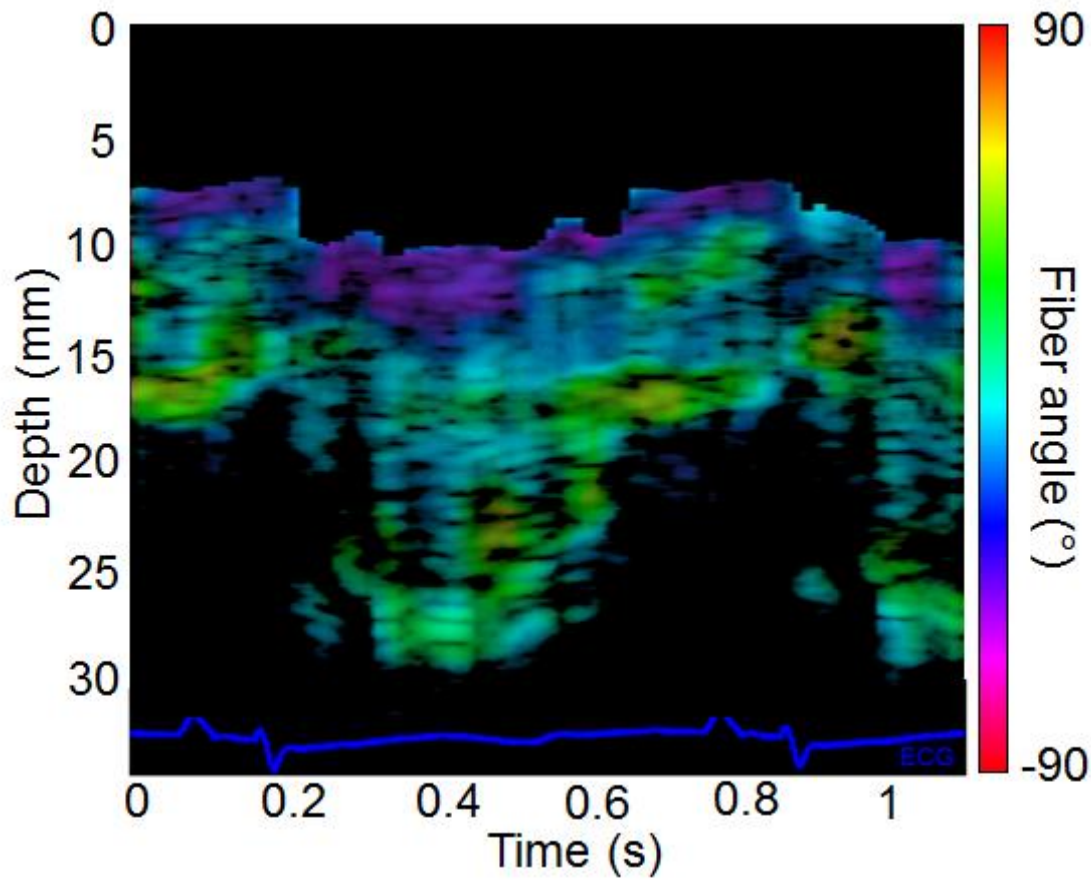

*figure 5*      ***Fiber orientation temporal variation***. *In vivo fiber angle variation in the left ventricular wall within a cardiac cycle as a function of depth and time. The color scale represents the value of the fiber angle superimposed onto a standard M-mode of the left ventricle. The associated ECG is displayed at the bottom of the figure.*

*In vivo transthoracic BTI of the human heart*

Finally, the feasibility of 3D-BTI on the human heart was shown on a healthy volunteer in a realistic clinical transthoracic imaging setting. Transthoracic conventional gray-scale imaging was used to position the probe on the parasternal view. 3D-BTI was then performed with a total imaging depth of 60 mm. Figure 6 shows the helicoidal fiber distribution imaged in the left ventricle anterior wall at mid-level both in systole and in diastole. Fiber angles were found to vary transmurally with an absolute difference of 104±9° from the epicardium to the endocardium.

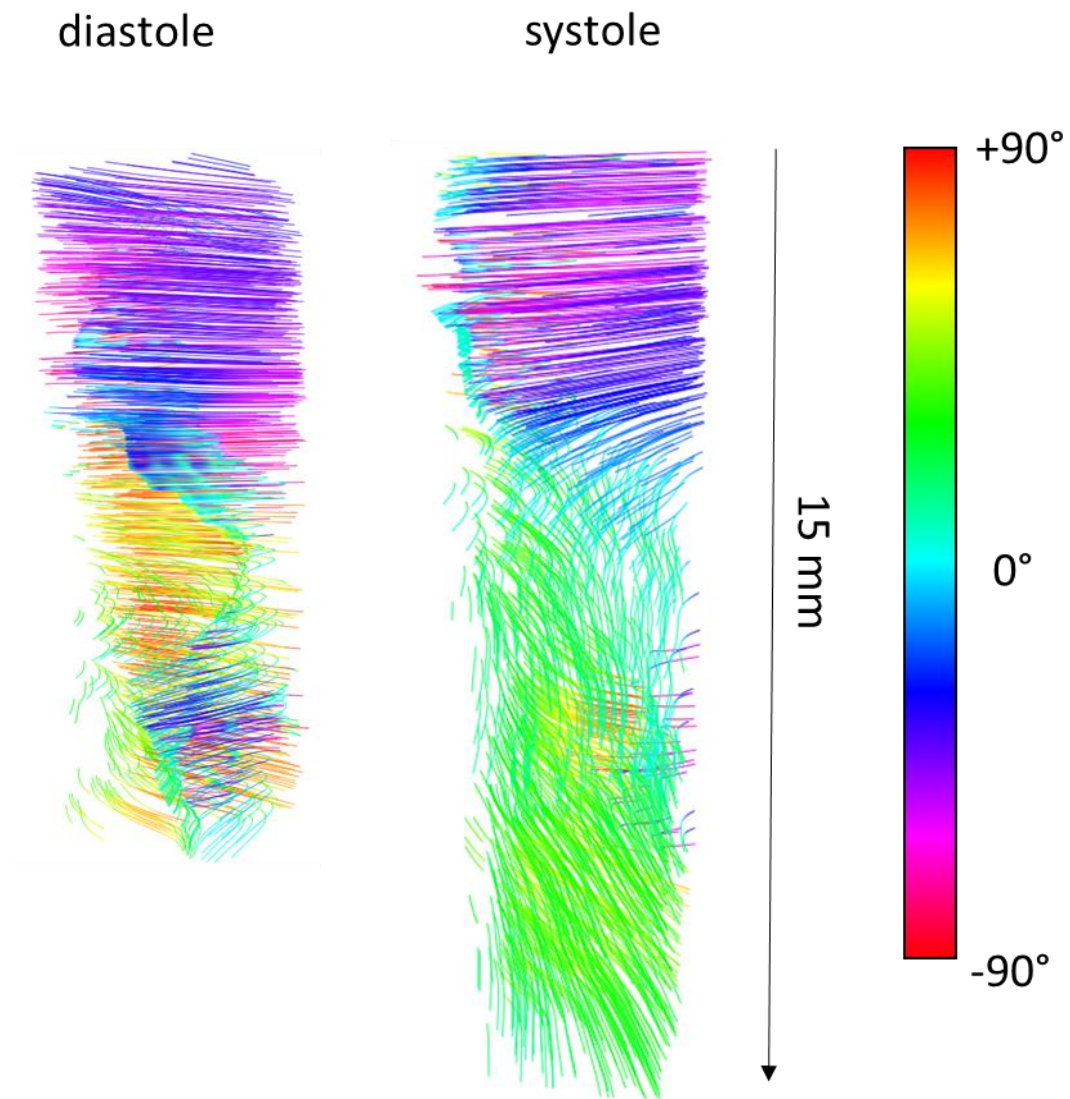

*figure 6*   ***Transthoracic imaging of myocardial fiber orientation in the human heart.***
*The fiber orientations of the antero-septal LV wall is shown in end-diastole and end-systole.*

**Discussion**

In this study, we have introduced 3D-BTI, a novel ultrasound-based technique for the mapping of the fibers orientation in myocardial tissues. The objectives were to determine the accuracy and precision of 3D-BTI in mapping the fibers structure and to demonstrate its feasibility on beating hearts. The accuracy and precision of 3D-BTI were validated in 5 *ex vivo* porcine left ventricular myocardial samples. The fibers angles distribution of each sample was obtained within volumes of 3 cm$^3$ and the transmural angle variations of each sample were compared to histology. An excellent agreement was found against histology, therefore demonstrating the accuracy and precision of 3D-BTI.

The in vivo feasibility of 3D-BTI was then demonstrated in a beating sheep heart. The fibers distribution of the left ventricular free wall was successfully imaged during one entire cardiac cycle at a volume rate of 90 volumes/s. To our knowledge, this is the first time that the dynamics of the myocardial fibers were mapped during an entire cardiac cycle at high frame rate. The transmural fiber variation was found to be 96° in diastole and did not change significantly during the systolic phase despite a 56% increase in wall thickness. This small variation over the cardiac cycle confirms the results of several studies that have investigated the fibers orientation in diastole and in systole using histology[27]. Finally, the feasibility of BTI in a realistic clinical transthoracic imaging setup was shown on a healthy volunteer and allowed for the mapping of the myocardial fibers orientation of the human heart both in systole and in diastole.

3D-BTI could not only bring new insights in the knowledge of the cardiac mechanics but it could also become a major tool to non-invasively detect fiber disorders linked to cardiac diseases in early stages. Myocardial fiber disarray has been shown in post-infarct remodeling of the left ventricle as well as in hypertrophic cardiomyopathy. 3D-BTI may be used to quantify disarray in the fiber angles but also to map the fractional anisotropy in order to quantify collagen infiltration and fibrosis content. Fractional anisotropy provides a measure of the level of anisotropy and can be obtained using the coherence values along and across the fibers.

3D-BTI presents a number of advantages as it can be performed non-invasively at high frame rate, and *in vivo* in beating hearts. In contrast to shear wave imaging based techniques, 3D-BTI is based on low energy emissions and thus can be performed continuously to provide the complete description of the fiber dynamics at high frame rate. Moreover, while the present study focused on the use of plane waves for the generation of synthetic foci, diverging waves[28,29], which can achieve even larger fields of view and image the entire heart in 3D[26], could also be used to improve 3D-BTI. The use of diverging waves would however result in a non-uniform spatial resolution so that this limitation needs to be evaluated in more details.

3D-BTI is robust to motion as a result of the use of ultrafast volumetric acquisitions, which enabled the mapping of the coherence functions in entire 3D volumes using a limited number of transmits. For example, while 49 plane wave emissions were sufficient to perform 3D-BTI in each voxel, performing the equivalent acquisition using standard focusing would have required 64*64*500 emissions, which would corresponds to a few minutes per volume instead of a few milliseconds in the case of plane wave imaging.

One of the current limitations of 3D-BTI is its incapability to map the z component of the fiber direction. Despite this limitation, mapping of fiber directions remains possible in anisotropic soft tissues in which the fibers are parallel to the 2D array plane, which is the case for the skeletal muscle or the anterior wall of the myocardium in standard parasternal short and long axis views. However, to apply 3D-BTI to other tissues in which structures can be oriented along the z-axis in sagittal and coronal views, tilted synthetic foci would be required. This could be achieved using subapertures in reception[30], which is the object of on-going work in our group. Moreover, while the ultrasound acquisitions required to perform 3D-BTI are very short (~10 ms), postprocessing operations required up to twenty minutes to achieve the vectorial representation of a full volume. Further optimization of the algorithms along with the rapid growth of computational power of video cards and processors could potentially lead to a real-time implementation of 3D-BTI in the future. Segmentation of the epicardium and endocardium remained also challenging using the Bmode images provided by the BTI acquisition and required to be manually performed by a trained cardiologist. A dedicated anatomical high quality imaging sequence based on harmonic imaging could improve considerably the detection of the wall boundaries.

BTI may be affected by low ultrasound signal to noise ratio at large depths. The performance of BTI need to be investigated in more details for low SNR configuration. We can anticipate BTI to be robust to incoherent noise such as clutter noise and thermal noise, thanks to the suppression of a large part of incoherent noises by the coherence function. However, we can also anticipate that the performance of BTI will drop considerably in very low SNR situations where incoherent noise becomes higher that the coherent signals.

Finally, while presenting similar advantages in terms of portability, safety, and real-time imaging capabilities, the 3D ultrafast ultrasound system that we used is a unique laboratory prototype that is not currently available in clinical practice. In summary, 3D-BTI may constitute a unique tool for non-invasively evaluating the myocardial fiber structure of a patient suffering from myocardial fibrosis or hypertrophic cardiomyopathy for diagnosis, treatment monitoring and follow-up.

**Materials and methods**

*Spatial coherence on a matrix array transducer*

Ultrasonic spatial coherence was assessed experimentally by focusing an ultrasound wave in a medium and by receiving the backscattered echoes on a 2D matrix array transducer (Figure 7.A and B). The 2D spatial coherence function $R(\Delta x, \Delta y)$ was obtained by computing the auto-correlation of the signals received by pairs of elements $i$ and $j$ distant by $\Delta x$ and $\Delta y$, along the main two coordinate axis of the matrix array:

$$R(\Delta x, \Delta y) = \frac{1}{N_x - |\Delta x|} \frac{1}{N_y - |\Delta y|} \sum_i \sum_j \frac{\sum_{t=T_1}^{T_2} S_i(x_i, y_i, t)\, S_j(x_j, y_j, t)}{\sqrt{\sum_{t=T_1}^{T_2} S_i(x_i, y_i, t)^2\, S_j(x_j, y_j, t)^2}} \quad (1)$$

Where $S_k(x_k, y_k, t)$ is the signal received on the element k of the matrix transducer with coordinates $x_k, y_k$ after applying a time-delay to compensate the propagation path. $\Delta x$ and $\Delta y$ are the distances between and element $i$ and element $j$:

$$\Delta x = x_i - x_j\,;\ \Delta y = y_i - y_j \quad (2)$$

$T_1$ and $T_2$ represents the averaging temporal window (equivalent to depth).

In random media, i.e. in absence of fibers, ultrasound scatterers are randomly distributed, (figure 7.A), so that the coherence function is predicted by the so-called Van Cittert-Zernike theorem[31]: the coherence function is given by the spatial Fourier transform of the intensity distribution at the focal spot. For a square matrix array, the pressure distribution at the focal spot is a 2D sinc function so that the coherence function appears as the Fourier Transform of a squared 2D sinc function. The pyramidal shape coherence function of a $N_x \times N_y$ 2D matrix array transducer is displayed on figure 7.C.

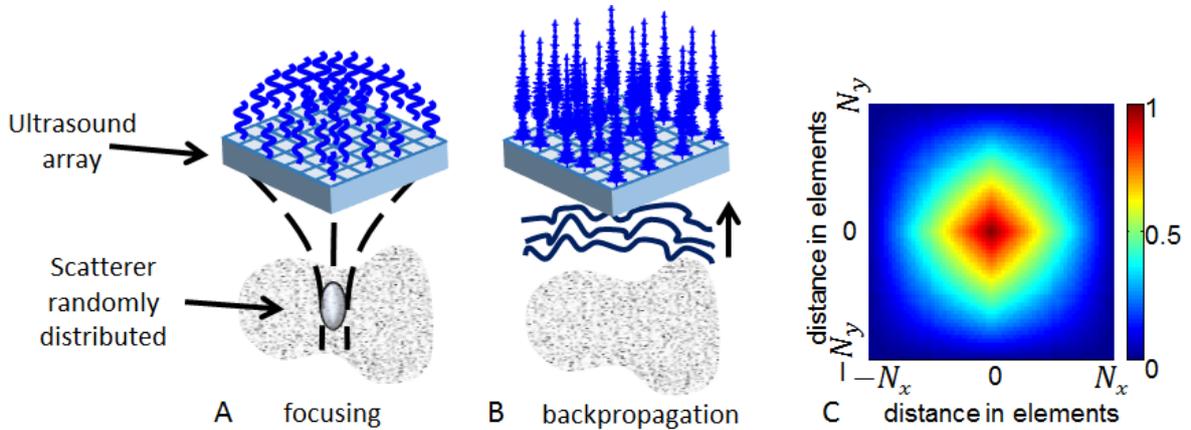

*figure 7* **Principle of spatial coherence estimation on a 2D matrix array probe**. *A An ultrasound wave is focused in a random media by applying delays in emission. B The backscattered echoes are recorded for each element of the 2D array. The autocorrelation of the rephased signals is calculated for each pair of elements as a function of distance between elements that composed the pair. C represents the theoretical coherence function for a focused emission in a random media that we would obtain with a matrix probe of $N_x \times N_y$ elements.*

*Data acquisition and signal processing*

In order to focus the ultrasonic wave at specific locations in the medium, conventional ultrasound imaging relies on applying time delays to the transmitted wave (figure 7). A large number of transmitted waves focused at different locations of the medium is then required to obtain one single image of the medium. In volumetric imaging, the number of transmitted waves can reach several thousands of waves which results in very low volume rates. To overcome this issue, our approach consisted in using ultrafast plane wave imaging with a coherent compounding approach[26]. Coherent compounding can generate, at any location of the region of interest, synthetic focal zones using only a few tens of emissions. Therefore, with this approach, large volumes can be imaged at very high volume rate between 50 and 5000 volumes/s.

The emissions consisted in the transmission of several tilted 2D plane waves (i.e. figure 11.A). Each plane wave was defined by two angles. All pairs of angles ranging from -6° to 6° with a step angle of 2° were emitted, for a total of 49 tilted plane waves. Plane waves were emitted from a 2D matrix array probe with 32 (x-axis) by 35 (y-axis) elements (3MHz, 0.3 mm pitch, 50% bandwidth at -3dB, Vermon) connected to a customized, programmable, 1024 channel ultrasound system described in Provost et al.[26]. This system was designed and built in-house for 3D ultrafast imaging applications. The 1024 independent channels could be used simultaneously in transmission, whereas receive channels were multiplexed to 1 of 2 transducer elements. Therefore, each emission was repeated twice, with the first half of the elements receiving during the first emission, and the second half of the elements receiving during the second emission.

The received signals were sampled at 12 MHz, for all the 1024 elements and the radio-frequency data were recorded. The 3D images were computed off-line using a dynamic receive focusing beamforming algorithm followed by coherent compounding (see figure 11.B). The 3D volume had a lateral size of 9.6x9.6 mm and a 3 cm to 6 cm depth depending on the region of interest. The lateral sampling of the image was 0.3 mm laterally (32x32 lines) and the axial sampling was 0.05 mm.

2D coherence functions were computed at each point of the 3D volume (figure 11.C) using equation (1). The temporal window $[T_1\ T_2]$, was set to 1.6 µs (i.e. five periods at 3 MHz). The coherence functions were averaged at each depth of the 3D volume using a lateral sliding window with a kernel of (0.15x0.15 mm).

An elliptic fit was applied on the averaged coherence functions on arbitrarily fixed 0.4 autocorrelation coefficient threshold. The direction of the main axis of the ellipse was used to define the direction of the fibers at each point of space (i.e. figure 11.D). Finally, a vector representation was used to visualize the fibers directions in 3D using the Amira software (Visualization Sciences Group, Burlington, MA) (i.e. figure 11.E).

*Experimental Setup*

1) In vitro experiments and histology

In vitro experiments were performed on five porcine fresh left ventricle myocardial samples embedded in a 2% gelatin gel. A cross-marker whose orientation matched the local cardiac circumferential-longitudinal coordinates was labeled on the myocardial sample. The probe was positioned at 15 mm of the epicardium. Each myocardial region examined by BTI was dissected from the intact porcine heart into a rectangular block (10x10x25 mm). This block was fixed in formalin for 48h. To keep the orientation of the tissue block in the wright direction, the basal aspect of the tissue block was labeled with tattoo ink. Then the tissue block was cut in 4 to 5 tissue slices which were embedded in paraffin. Each paraffin block was completely cut at 5 µm of thickness in serial sections. One histological section out of 50 was stained with H&E: the resulting gap between two stained adjacent sections was 250 µm. All the sections were scanned with a Hamamatsu Nano Zoomer (Hamamatsu City, Japan). Fiber angles in all the histological digitalized images studied were computed using the Hough transform[32].

2) In vivo experiments

Animal experiments were performed in an open chest sheep on the mid-anterior region. The experimental procedure was approved prior to use by the Institutional Animal Care and Use committee of Hôpital Européen Georges Pompidou (PARCC) according to the European Commission guiding principles (2010/63/EU). 3D-BTI was thus performed within an entire cardiac cycle. The 3D-BTI acquisition was performed at a volume rate of 90 volumes/s during 1.2 seconds in order to cover more than one cardiac cycle. It enabled the reconstruction of 100 fiber volumes. During the experiment, the electrocardiogram was recorded.

Human experiments were performed by a trained cardiologist. The volunteer was positioned on the left lateral decubitus position. The probe was placed in the parasternal view and 2D real-time imaging was performed to position the probe and image the antero-septal wall. The 3D BTI acquisition was then launch to acquire 1s of data at a rate 90 volumes/s. The study received the approval of the CPP (Paris Ile de France VI, N° 15-15).

**Acknowledgements**

This work was supported by the European Research Council under the European Union's Seventh Framework Program (FP/2007–2013) / ERC Grant Agreement n°311025 and by LABEX WIFI (Laboratory of Excellence ANR-10-LABX-24) within the French Program "Investments for the Future" under reference ANR-10-IDEX-0001-02 PSL. J.P. was funded in part by a Marie Curie International Incoming Fellowship under the Seventh Framework Program for research (FP7) of the European Union.

**Author contribution statement**

C.P, J.P, O.V and M.P wrote the main manuscript. C.P, V.F, J.P, J.L.G, O.V and M.P performed experiments. P.B. performed histology. M.T, M.F and M.P designed study. All authors reviewed the manuscript.

**Additional Information**

Competing financial interests : The authors declare no competing financial interests.

**Figure Legends**

figure 1    **3D-BTI principle**. *A. Tilted plane waves are emitted and the associated backscattered echoes are recorded. B. Coherent compounding is applied in postprocessing to synthetically generate voxel-specific focal zones. C. The 2D spatial coherence function is calculated for each voxel and D. an elliptic fit is applied to determine the fibers orientation. E. a vector representation is then achieved to display the fibers orientation in 3D.*

figure 2    **3D-BTI of one ex vivo porcine myocardial sample**. *A 3D grayscale ultrasound image of one myocardial sample. B, C, D Three examples of spatial coherence functions obtained for different depth locations in the sample. E drawing of the heart associated to the ultrasound acquisition location. F Vector representation used for the visualization of the transmural variation of the fibers in the entire sample. G* ***Schematic representation of the reference coordinates.*** *Local axes u, v and w are defined by the circumferential and longitudinal directions on the epicardial surface. Fiber angle α, is positive when measured counterclockwise from the v-axis. (Figure adapted from Streeter et al. [27])*

figure 3    **3D-BTI validation against histology.** *A. Five histological slices (20X) and associated BTI slices of a left ventricle myocardial sample highlights the transmural angle variation of the fibers. B. An example of the fiber angles variation assessed with histology and with 3D-BTI through the wall thickness (0% epicardium-100% endocardium) for the same sample is displayed. C. Bland-Altman plot of the transmural fiber angles estimated by 3D-BTI and histology (n=5 myocardium samples).*

*figure 4* **In vivo 3D-BTI.** 3D representation of fibers orientation in the left ventricle of an open-chest sheep at four different moments of the cardiac cycle (i.e. A Late Diastole, B Early Systole, C Late Systole, D Early Diastole).

*figure 5* **Fiber orientation temporal variation**. In vivo fiber angle variation in the left ventricular wall within a cardiac cycle as a function of depth and time. The color scale represents the value of the fiber angle superimposed onto a standard M-mode of the left ventricle. The associated ECG is displayed at the bottom of the figure.

*figure 6* **Transthoracic imaging of myocardial fiber orientation in the human heart.** The fiber orientations of the antero-septal LV wall is shown in end-diastole and end-systole.

*figure 7* **Principle of spatial coherence estimation on a 2D matrix array probe.** A An ultrasound wave is focused in a random media by applying delays in emission. B The backscattered echoes are recorded for each element of the 2D array. The autocorrelation of the rephased signals is calculated for each pair of elements as a function of distance between elements that composed the pair. C represents the theoretical coherence function for a focused emission in a random media that we would obtain with a matrix probe of $N_x \times N_y$ elements.